\documentstyle[aps,twocolumn,epsf]{revtex}
\begin{document}
\twocolumn[
\hsize\textwidth\columnwidth\hsize\csname@twocolumnfalse\endcsname
\draft

\title{Heat Conduction in $\kappa$-(BEDT-TTF)$_2$Cu(NCS)$_2$}
\author{St\'ephane Belin, Kamran Behnia}
\address{Laboratoire de Physique des Solides(CNRS), Universit\'{e} Paris-Sud,
F-91405 Orsay, France}
\author{ Andr\'e Deluzet}
\address{Institut des Materiaux de Nantes, Universit\'{e} de Nantes, 
F-44322 Nantes, France}
\date{May 27, 1998}

\maketitle

\begin{abstract}
The first study of thermal conductivity, $\kappa$, in a quasi-two-dimensional 
organic superconductor of the $\kappa$-(BEDT-TTF)$_2$X family reveals 
features analogous to those already observed in the cuprates. The 
onset of superconductivity is associated with a sudden increase in 
$\kappa$ which can be suppressed by the application of a moderate magnetic
field. At low temperatures, a finite linear term - due to a residual 
electronic contribution- was resolved. The magnitude of this term
is close to what is predicted by the theory of transport in unconventional 
superconductors.

\end{abstract}
\pacs{}]


The superconductors of $\kappa $-(BEDT-TTF)$_2$X family\cite{lang} share a 
number of similarities with the high-T$_c$ cuprates\cite{mckenzie}. Both 
sets of compounds are quasi-two-dimensional with superconductivity confined
to conducting planes sandwiched between insulating layers. The metallic 
state in both families exhibit common features like low carrier densities, strong 
electronic correlations and proximity of antiferromagnetic insulating 
state. While Shubnikov-de Haas experiments\cite{caulfield}
have established the existence of a well-defined Fermi surface in the 
$\kappa $-(BEDT-TTF)$_2$X family, this metallic state presents 
some more unconventional properties - like a  pseudogap in the electronic 
density of states in $\kappa $-(BEDT-TTF)$_2$Cu[N(CN)$_2$]Br\cite{kawamoto}-
which have been compared to analagous features in underdoped 
cuprates\cite{mckenzie}. As for the symmetry of the superconducting order 
parameter, it has yet to become the subject of a consensus as nowadays it is 
the case in the cuprates. While, early penetration-depth studies on 
$\kappa$-(BEDT-TTF)$_2$Cu(NCS)$_2$ led to conflicting results \cite{penetration},
recent NMR\cite{mayaffre} and specific heat\cite{nakazawa} studies on 
$\kappa $-(BEDT-TTF)$_2$Cu[N(CN)$_2$]Br provided evidence for the presence of 
nodes in the superconducting gap. 

In this letter we present the first study of 
thermal conductivity in a member of this family. 
According to our results, heat transport in $\kappa$-(BEDT-TTF)$_2$Cu(NCS)$_2$ 
presents features which have  already been detected in YBCO and other high-T$_c$ cuprates. 
Notably, the observation of a residual electronic thermal conductivity 
at very low temperatures provides strong support for presence of nodes 
in the superconducting order parameter. 

We measured the thermal conductivity of five $\kappa $-(BEDT-TTF)$_2
$Cu(NCS)$_2$ single crystals  using a conventional four-probe method. 
Contacts were realized using silver paint on evaporated gold. The heat current was 
always applied in the basal (highly-conducting) plane. The temperature gradient was 
measured with two  RuO$_2$ resistance chips which showed small magnetoresistance and 
a usable sensitivity up to 15K. The resistive heater and the two thermometers were held 
by small solenoids of 50 $\mu $m manganin wire. In this way, we measured
the resistance of the sample and the thermometers with minimal heat loss.
For temperatures below 0.25 K, we checked our zero-field results by using another
device which was designed for very low temperatures and described 
elsewhere\cite{belin}. Our set-up allowed us to measure, in addition to 
electrical and thermal conductivities, the thermo-electric power of the sample. Direct 
visual measurements of sample dimensions led to a
gross determination of the geometric factor($\pm 50 \% $) due to 
irregularities in samples' shape and thickness. Determining the absolute
value of the resistivity proved to be very difficult.  We found room temperature 
resistivities varying from 35 to 80m$\Omega$ cm. A comparable dispersion can be
found in the technical litterature on this compound. We used a unique 
room-temparature-resistivity  value (54 m$\Omega$ cm) when comparing different samples. 
At very low temperatures, we found that in all samples
the voltage signal of the standard IVVI configuration was less than the IIVV one 
(the order of Is and Vs refer to the spatial sequence of current(I) and voltage(V) 
electrodes on the sample). This is a signature of a meandering charge current path
 caracteristic of highly-anisotropic superconductors with inhomogenous contacts
\cite{aukkaravittayapun}. Therefore, we refrained to use  the nominal value of 
 resistivity in our analysis of thermal  conductivity data. 
 
Fig. 1 presents the effect of the superconducting transition on the 
temperature dependence of the thermal conductivity. We present the
results for the two samples which were most thoroughly
studied. The striking feature of the figure is the upturn in 
thermal conductivity at the onset of the superconducting transition. All 
the samples studied presented such an upturn, but its intensity -reflected 
in the height of the consequent peak in $\kappa$(T)- was found to be 
strongly sample-dependent. The ratio $\frac{\kappa_{max}}{\kappa(T_c)} $ is
1.3 in sample \#1 and  1.05 in sample \#2. We observed a ratio as high as
2.7 in one sample which was quickly deteriorated after a thermal cycle.
 This strong variation suggests that heat conduction is much sensitive to 
a type of disorder which affects only mildly the electrical resistivity. 
The residual restivity ratio $\frac{\rho(300K)}{\rho(T \rightarrow 0)}$ is
410 in sample \#1 and 250 in sample \#2. The positive sign of thermopower for both 
samples indicated that the orientation of heat current was nearly parallel to the 
quasi-one-dimensional sheets of Fermi surface and thus mainly 
implying the hole-like carriers of the two-dimensional pockets\cite{logenov}. 
 \begin{figure}[tbph]
\epsfxsize=8.5cm
 $$\epsffile{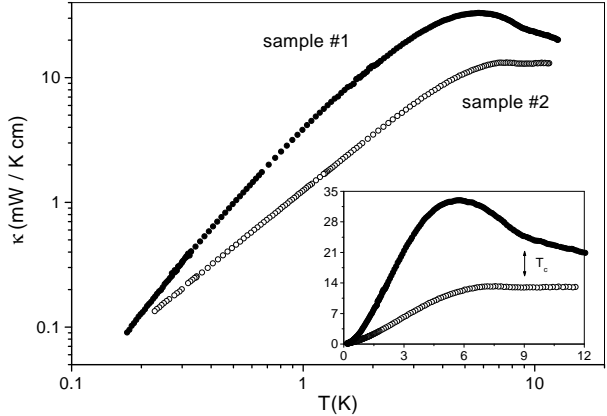}$$
\caption{Temperature dependence thermal conductivity in two different
 samples. Note the upturn at T$_c$. Inset shows a linear presentaion.}
\label{fig1}
\end{figure}
 A similar upturn in the thermal conductivity of high-T$_c$ superconductors
has been a subject of controversy for several years\cite{yu}. The increase in 
thermal conductivity below T$_c$ indicates that the condensation of 
electrons in the superconducting state strengthens heat transport by 
reducing the scattering of heat carriers. The debate was centered on the 
identity of these heat carriers. While an orthodox scenario\cite{cohn} invoked an 
increase in the lattice conductivity due to condensation of electrons, 
experimental evidence for a very unusual increase in the electronic 
relaxation time in the superconducting state\cite{bonn} led to the 
suggestion\cite{yu} that 
at least part of the upturn in $\kappa $ is due to a steep increase in the 
electronic contribution. Strong support for the latter point of view was 
provided by thermal Hall effet measurements\cite{krishana}.
In the case of $\kappa$-(BEDT-TTF)$_2$Cu(NCS)$_2$, the origin of the 
upturn raises the same questions. Surface resistance studies have 
reported an increase in the microwave conductivity of the system below 
T$_c$\cite{achkir}. Compared to YBCO\cite{bonn} this increase is 
modest, but its very existence makes it tempting to stretch the analogy 
with the cuprates and suggest that part of the upturn in $\kappa$(T) 
reported here is due to electrons. However, the Wiedmann-Franz law (with 
should be employed very cautiously due to the uncertainties on the absolute value 
of resistivity) implies that just above T$_c$, the 
electronic contribution counts for only 5\% of the
total thermal conductivity. Thus, while the final issue of the question 
shall wait for thermal Hall effect measurements in the superconducting 
state of this compound, one can safely attribute the main part of this 
feature to the enhancement in the lattice conductivity consequent to a 
sudden decrease in electronic scattering at T$_c$. This very visible effect
of electronic condenstation on lattice conductivity indicates the strength
of the electron-phonon coupling in this system as already documented by 
neutron diffraction \cite{pintschovius} and Raman scattering\cite {pedron} 
studies. This is to be contrasted to the case of (TMTSF)$_2$ClO$_4$, where 
lattice conductivity was found to remain unchanged by the superconducting 
transition\cite{belin}. 
\begin{figure}[tbph]
\epsfxsize=8.5cm
$$\epsffile{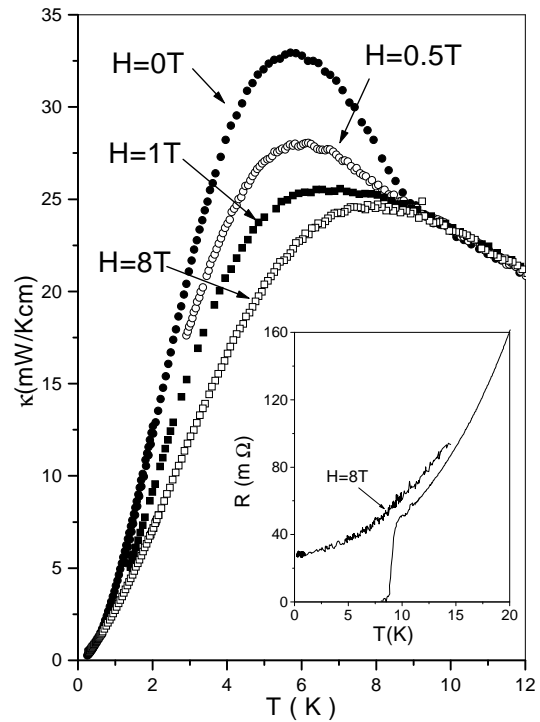}$$
\caption{The temperature dependence of the thermal conductivity in
sample\#1 for different fields . Inset shows the
temperature dependence of the resistance of the same sample.}
\label{fig2}
\end{figure}
A supplementary source of information is the effect of the magnetic field. 
Figure 2 shows $\kappa $(T) of sample\#1 for different values of 
magnetic field. The inset of the figure shows the electrical resisitivity 
of the normal and the superconducting states. In the normal state,
a magnetic field of 8T does not affect the thermal conductivity within the 
experimental resolution (< $ 1 \% $), but it induces
a sizeable (15 $\%$ ) decrease in charge conductivity. This is an 
additional indication of  lattice-dominated thermal conductivity in the 
vicinity of T$_c$. 
\begin{figure}[tbph]
\epsfxsize=8.5cm
$$\epsffile{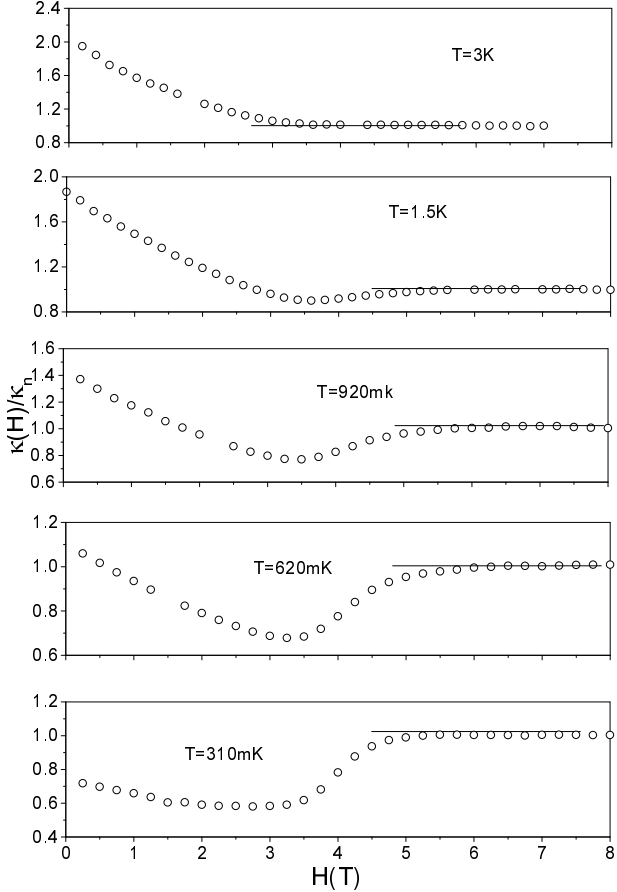}$$
\caption{The field dependence of the normalized thermal conductivity in 
sample\#1 for different temperatures. Deviation from the horizontal line
marks  H$_{c2}$. Note the gradual apparition of a  dip at low temperatures.}
\label{fig3}
\end{figure}
The peak is suppressed with the application of a moderate 
magnetic field. But the decrease in thermal conductivity is only monotonous at 
higher temperatures. This is seen in fig. 3 which presents the field-dependence
of thermal conductivity for various temperatures. To reduce undesired effects 
associated with vortex pinning, the sample was cooled in the normal state 
(i.e. in a field greater than H$_{c2}$) to the corresponding temperature 
and then $\kappa $ was measured as a function of decreasing field. For 
temperatures higher than 2K, thermal conductity decreases with increasing 
field as a result of the reintroduction of the scattering quasi-particles by 
the magnetic field. A more remarkable structure-reminiscent of the 
heavy-fermion superconductor URu$_2$Si$_2$\cite{behnia}- appears at lower 
temperatures when $\kappa $(H) exhibits a dip. This minimum indicates a 
competition between increasing and decreasing contributions to $\kappa $. Note 
that only the electronic component can be enhanced by the application of a 
magnetic field. Thus, the size of the jump in $\kappa $(H) just below H$_{c2}$ 
is an upper limit to the difference between electronic thermal conductivities in 
the normal and superconducting states. According to an early theory\cite{maki}, 
in the vicinity of H$_{c2}$, the fading of a spatially inhomogenous gap leads to 
a rapid enhancement in the density of states of  quasi-particles travelling 
perpendicular to the vortex axes. The slope of $\kappa $(H) at H$_{c2}$ is 
related to the toplogical details of the superconducting gap. 

 The main part of the 
initial field-induced decrease of thermal conductivity
is due to the effect of the magnetic field on the phonon mean-free-path. However,
 an estimation of the dominant phonon wavelength at low temperatures
($\lambda_{ph} = \frac{\hbar v_s}{k_B T} \simeq  240 nm/K $) exceeds by  two
order of magnitudes the coherence length at T=0.62K  so that no vortex scattering
of phonons is expected. This is confirmed by the regular decrease in $\kappa $(H)
up to fields of a few teslas. The vortex scattering of heat carriers 
has been reported in much smaller fields with long intervortex 
distances\cite{redko}. Here, the observed field-induced decrease is a result 
of the scattering of phonons by electronic excitations including those which are 
extended out of the vortex cores. As first pointed out by Volovik\cite{volovik},
the enhancement of these latter delocalized electronic excitations by a magneic 
field due to a doppler shift in quasi-particle spectrum dominates the properties 
of the mixed state of unconventional superconductors\cite{kubert}.  

Evidence for unconventional superconductivity comes from our low-temperature
results. Fig. 4 presents the low-temperature behaviour of thermal conductivity 
in normal and superconducting states for samples \#1 and \#2. 
The remarkable feature of the figure is the presence of a finite linear 
term in the thermal conductivity of the superconducting state indicative of a 
residual electronic  contribution. In both samples the magnitude of this term 
is a sizeable fraction of the normal electronic term. For sample \#1, we 
extended our zero-field measurements down to T=0.16 K and found that for T< 0.27, 
$\kappa$ (T) presents a $aT+ bT^3$ temperature dependence with  
$a=\kappa^s_e/T 0.20\pm0.09mW/K^2cm$ and $b= 11 \pm5mW/K^4cm$. 
The large uncertainties are mainly due to the geometric factor. The cubic term
gives an estimation of the maximum phonon mean-free-path using the kinetics
gaz equation  $\kappa _{ph}=\frac 13c_{ph}v_sl_{ph}$, where $c_{ph}=\beta T^3$
is the lattice specific heat ($\beta = 23.6 \mu J/K^4cm^3$\cite{goodrich})
and v$_s$ the velocity of sound (v$_s$= 5.10$^3$m/s\cite{yoshizawa}). This
yields $l_{ph}$= 28$\mu m$ which is comparable to the sample 
tickness($\simeq $ 20$\mu m$). In the normal state, due to the lack of 
data for T < 0.25k the extraction of  $\kappa^n_e$/T  value at T=0 is
less straightforward. But one can reasonably expect that the ballistic
regime (with a phonon mean-free-path comparable to sample dimensions)
should be attained at a similar temperature range. Moreover, the magnitude
of the cubic term( which depends on phonon thermodynamics and sample
geometry) should be identical in the normal and superconducting states.
In this way, the zero-temperature $\kappa^n_e$/T can be estimated 
to be 0.95 $ mW/K^2cm$. As expected, the difference 
between the electronic thermal conductivities of the normal and superconducting
states is comparable with the jump in $\kappa(H)$/T  just below  H$_{c2}$ at 
T= 0.31 K, which- as argued above- is exclusivlely electronic and gives an estimate 
of $\kappa^n_e - \kappa^s_e$. In the case of sample \#2, due to the lack of
low-temperature data, the analysis remains qualitative. 

The theory of heat transport in unconventional superconductors predicts a finite
zero-temperature value for $ \frac{\kappa _s}{T} $ due to impurity scattering of 
residual quasi-particles\cite{graf}. Moreover, for certain gap topologies- 
including the one associated with the d$_{x^2-y^2}$ symmetry- the magnitude of this 
linear term is universal at small concentrations of impurity:
$\frac{\kappa _{00}}{T}= \frac{\hbar k_b^2 \omega_p^2}{6e^2S}$, where $\omega _p$ 
is the the plasma frequencey and  $ S=\frac{d\Delta_0}{d\Phi} $ is the slope of the
gap at the node.  
The experimental validity of this theory has been 
recently reported in the case of YBa$_2$Cu$_3$O$_{6.9}$\cite{taillefer}.
In our case, using $\hbar \omega_p\simeq 0.6\pm 0.1eV$\cite{ugawa} and  
2$\Delta _0 = 4.8 \pm 1.1meV$\cite{bando} and assuming a standard d-wave gap with
$ S= 2 \Delta_0 $  one expects  $\frac{\kappa _{00}}{T} = 0.16 \pm0.05 mW/K^2cm$
in very good agreement with the experimental result.

\begin{figure}[tbph]
\epsfxsize=8.5cm
$$\epsffile{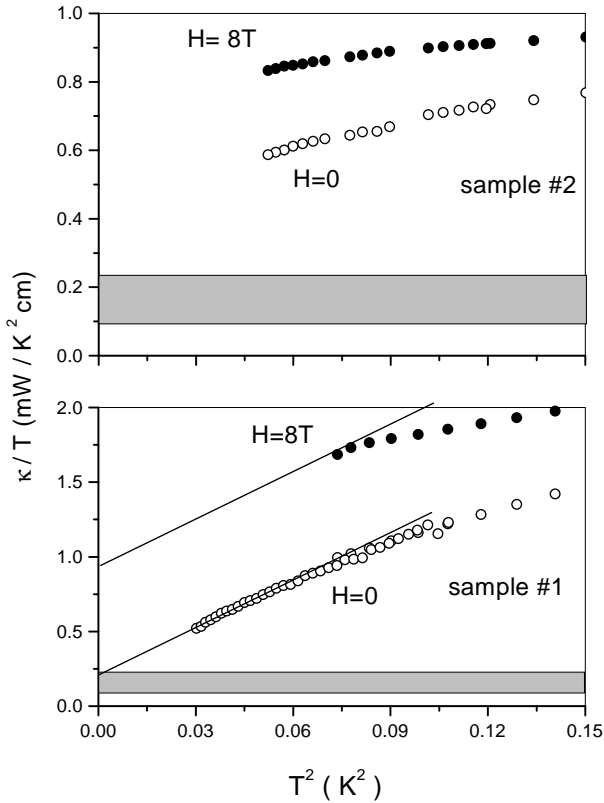}$$
\caption{The $T^2 $ dependence of $\kappa /T$ in normal
and superconducting states for sample \#1  and sample \#2. The 
straight lines schematize the expected $aT + bT^3 $ behavior. The shaded area 
represents the universal linear term  in the clean limit.}
\label{fig4}
\end{figure} 
It is important to note that this theory has been worked out for clean 
superconductors with the electronic relaxation time, $\tau $, exceeding  
$\frac{\hbar}{\pi \Delta_0} $. An alternative 
formulation  of its basic statement is  that the ratio of electronic
conductivities in the superconducting and normal states should extrapolate
to $\frac{\hbar}{\pi \tau \Delta_0} = \frac{\xi_0}{l_e} $. Compared to
YBCO, $\kappa $-(BEDT-TTF)$_2$Cu(NCS)$_2$ is a rather dirty superconductor.
Its electronic mean-free-path , $ l_e $, can be estimated either using
the magnitude of the threshold field for apparition of quantum 
oscillations ( 8T)\cite{caulfield} or through the values of normal state
conductivity ( $\kappa^n_e$/T ) and $\omega_p $.  Both methods 
yield comparable values for $ l_e $ ( $ \simeq$  35nm  ). The coherence length 
$ \xi_0 $  can be deduced from the slope of H$_{c2}$ at T$_c$ ( $ \simeq$ 5 nm). 
Thus the value of $ \frac{\kappa^s_e}{\kappa^n_e}$ should extrapolate to $0.15 \pm 0.05 $.
According to theory, this value should increase with increasing 
scattering rate\cite{sun}, so that for dirtier samples higher ratios are expected. 
Fig. 4 is compatible with both of these statements. While for sample \#1 we have a remarkable
quantitative agreement with the theory, the extrapolated ratio in sample \#2 tends
to be higher. This is the first verification of the theory of transport in 
unconventional superconducors by directly comparing the conductivities of normal 
and superconducting states. 

In conclusion, our study of heat conductivity in $\kappa $-(BEDT-TTF)$_2$Cu(NCS)$_2$ 
provides evidence for nodes in the superconducting gap, strong electron-phonon 
coupling and possibly an enhancement of quasi-particle scattering time below 
T$_c$. We thank H.Aubin, L. Taillefer, M. Ribault, C. Pasquier, D. J\'erome and 
L. Fruchter for stimulating discussions, P. Batail for his support and
L. Bouvot for technical assistance.

\end{document}